\title{Asymptotic Dynamics of Monopole Walls}
\author{
        R. Cross \\
                \textit{Department of Physics, University of Arizona, Tucson, AZ 85721, USA} \\
                rsc42@physics.arizona.edu
}

\documentclass[10pt,fleqn]{article}
\usepackage{cancel,cite,footnote,footmisc,pdfpages}
\usepackage{amsmath, amsthm, amssymb,bbold,slashed}
\usepackage{graphics,graphicx,bbm}
\usepackage[margin=1.5in]{geometry}
\usepackage{hyperref}

\hypersetup{
    colorlinks,
    citecolor=black,
    filecolor=black,
    linkcolor=black,
    urlcolor=black
}

\begin{document}
\maketitle

\textbf{Abstract}

We determine the asymptotic dynamics of the U(N) doubly periodic BPS monopole in Yang-Mills-Higgs theory, called a monopole wall, by exploring its Higgs curve using the Newton polytope and amoeba. In particular, we show that the monopole wall splits into subwalls when any of its moduli become large. The long-distance gauge and Higgs field interactions of these subwalls are abelian, allowing us to derive an asymptotic metric for the monopole wall moduli space.

\section{Introduction}

In 1931 \cite{Dirac:1931}, Dirac proposed a magnetic cousin to the electron in classical electromagnetism, now referred to as the Dirac magnetic monopole. Analogous to  the classical electron, it is a point particle and its magnetic field is singular. Nearly five decades later, 't Hooft and Polyakov \cite{'tHooft:1974qc,Polyakov:1974ek} expanded the idea of the magnetic monopole by identifying non-singular solutions now called 't Hooft-Polyakov monopoles in nonabelian Yang-Mills-Higgs theory, in which the Yang-Mills gauge fields couple to a scalar field with the usual symmetry-breaking Higgs potential. Prasad and Sommerfield \cite{Prasad:1975kr} found an explicit static $SU(2)$ solution for this theory in the massless Higgs limit. In the time-independent and massless Higgs limits, Bogmolny \cite{Bogomolny:1975de} derived his eponymous equation. Solutions to the Bogomolny equation solve the Yang-Mills-Higgs field equation and minimize energy. They are called BPS (Bogomolny-Prasad-Sommerfield) monopoles.  

Nonabelian magnetic monopoles are interesting in their own right, appearing as they do in many contestant grand unified field theories. They have garnered attention in recent decades, however, for their significance in relation to certain supersymmetric Yang-Mills quantum field theories. The nontrivial connection to these theories is via their moduli spaces of vacua. The moduli space of BPS Yang-Mills-Higgs monopoles (a set of solutions that share fixed boundary conditions and which together form a manifold) is isomorphic to the Coulomb branch moduli space of vacua in the associated super Yang-Mills theory \cite{Seiberg:1996nz,Chalmers:1996xh,Hanany:1996ie}. These moduli spaces are Calabi-Yau, specifically hyperk\"ahler, i.e. they are k\"ahler manifolds which are holomorphically symplectic.

In early studies of BPS monopoles, their moduli spaces were used to determine monopole behavior. Manton established  \cite{Manton:1985hs} that the low-energy dynamics for BPS monopoles can be approximated as geodesic motion on their moduli space. In the modern context, monopole moduli spaces have applications in quantum theories. Despite their importance, few metrics on monopole moduli spaces are known. BPS solutions in which some or all of the constituent monopoles are closely spaced represent points in the interior of the moduli space. BPS solutions in which the monopoles are very widely-spaced are points on the moduli space in its \textit{asymptotic region}. Long-range abelian approximations have been used to obtain the latter type of solution and metrics have been calculated for the corresponding asymptotic moduli spaces, but solutions of the former type have been mostly illusive. Because of this, most moduli space metrics that have been produced are accurate only for the asymptotic portion of the moduli space. The following paragraph enumerates these efforts. 

Atiyah and Hitchin \cite{Atiyah:1985dv} derived a metric on the \textit{full moduli space} for two $SU(2)$ BPS monopoles on $\mathbb{R}^3$. Gibbons and Manton \cite{Gibbons:1995yw} then generalized to $n$ BPS, \textit{well-separated}, indistinguishable $SU(2)$ monopoles and found the asymptotic moduli space metric. Lee, Weinberg, and Yi derived a similar asymptotic metric for general gauge symmetry \cite{Lee:1996kz}. Cherkis and Kapustin \cite{Cherkis:2000cj} used an approach echoing Gibbons and Manton's to determine the asymptotic moduli space metric for an $SU(2)$ monopole on $\mathbb{R}^2 \times S^1$ with $n$ indistinguishable charges, as did Hamanaka, Kanno, and Muranaka \cite{Hamanaka:2013lza} for an $SU(2)$ monopole on $\mathbb{R} \times T^2$ with $n$ indistinguishable charges. As mentioned, these monopoles arise in classical Yang-Mills-Higgs theory. Their moduli spaces are argued to be isometric to moduli spaces of vacua for $SU(n)$ super Yang-Mills quantum gauge theories with boundary conditions and dimension particular to each of the monopole periodicity cases. 
Seiberg and Witten originally discovered the existence of these relationships in \cite{Seiberg:1996nz}, following work by Seiberg and Witten \cite{Seiberg:1994rs, Seiberg:1994aj}, and Intriligator and Seiberg \cite{Intriligator:1994sm,Intriligator:1995id}. Chalmers, Hanany, and Witten \cite{Chalmers:1996xh, Hanany:1996ie} explained these relationships using brane dualities. Later Haghighat and Vandoren \cite{Haghighat:2011xx} examined the compacitified five dimensional quantum field theory relevant to doubly periodic BPS monopoles, and the underlying theory connecting them.

For $n$ monopoles on $\mathbb{R}^3$, this theory is related via the relative moduli space metric to the Coulomb branch $N=4$ $SU(n)$ super Yang-Mills vacuum in three dimensions \cite{Seiberg:1996nz}. For two such monopoles, the relative metric is called the Atiyah-Hitchin metric. $n$ periodic monopoles (on $\mathbb{R}^2 \times S^1$, called "monopole chains") are related via their moduli space metric to the Coulomb branch of vacua for $N=2$ $SU(n)$ super Yang-Mills in four dimensions which has been compactified on a circle \cite{Cherkis:2001gm}. Similarly, $n$ doubly-periodic monopoles (on $\mathbb{R} \times T^2$, called "monopole walls" or "monowalls") are related via their moduli space metric to the Coulomb branch of vacua for $N=1$ $SU(n)$ super Yang-Mills in five dimensions which has been compactified on a two-torus \cite{Cherkis:2012qs, Hamanaka:2013lza, Haghighat:2011xx}. 

This paper continues and elaborates on the efforts listed. Section $\ref{2}$ reviews Yang-Mills-Higgs theory, outlines our objectives, and reviews the Higgs spectral curve with its Newton polygon and amoeba. 
Section $\ref{3}$ begins with a BPS solution to the Bogomolny equation on $\mathbb{R} \times T^2$, a monowall, and shows that if it possesses moduli (degrees of freedom) then whenever a modulus becomes large the monowall can be interpreted as a collection of constituent monowalls which spread apart and become distinct, and their Higgs and electromagnetic interactions are abelian. In the case of the singly-periodic monopole, as in \cite{Cherkis:2001gm}, the Nahm transform maps the monopole onto a solution of the Nahm equations \cite{Nahm:1981nb}, formulating the problem of interacting monopoles as a Nahm system and validating the abelian approximation in the asymptotic regime. This approach is unsuccessful in the case of the doubly-periodic monopole, which is mapped to another doubly-periodic monopole under the Nahm transform. Instead we study some key behaviors of the doubly-periodic monopole using the Higgs spectral curve \cite{Hitchin:1982gh, Cherkis:2000cj, Cherkis:2000ft, Cherkis:2012qs}, which allows a geometrical treatment of the monopole interactions in the BPS limit. We demonstrate that if \textit{any} monowall has moduli then taking a modulus to infinity causes the monowall to break into subwalls. We model the asymptotic behavior of a general monowall as abelian interactions among its well-separated subwalls. Section $\ref{4}$ generalizes the monopole of \cite{Cherkis:2012qs, Hamanaka:2013lza} from $SU(2)$ to $U(N)$ for distinguishable subwalls. By modeling the interactions of well-separated nonabelian subwalls as the interactions of abelian monowalls, we determine an expression for a hyperk\"ahler asymptotic metric for subwalls widely-spaced in the non-compact dimension relative to the width of a single subwall, and discuss the symmetries of the metric. Our approach allows for subwalls which are Dirac monowalls (singularities), which have no dynamics of their own, but whose fields affect the motion of the remaining monowalls.

\section{Background and Setup} \label{2}

\subsection{Yang-Mills-Higgs Theory}

In classical, 3+1-dimensional $U(N)$ Yang-Mills-Higgs theory the pure Yang-Mills action is augmented by that of a scalar with the usual symmetry-breaking  potential. 
\begin{equation}
S = \int d^4x \text{Tr} \left[ \frac{1}{2} F_{\mu \nu} F^{\mu \nu} - \left( D_{\mu} \phi \right) \left( D^{\mu} \phi \right) - \lambda \left( \phi^2 + v^2 \right)^2 \right].
\end{equation}
We shall have both the gauge and Higgs fields antihermitian in the adjoint representation. They can be expressed as linear combinations of the antihermitian $U(N)$ generators $T_b$: $\phi = \phi^b T_b$, $A_{\mu} = A_{\mu}^b T_b$ where $b=1,...,N^2$, and $v$ is real. The gauge covariant derivative is $D_{\mu} \phi = \partial_{\mu} \phi + [A_{\mu},\phi]$ and the field strength is $F_{\mu \nu} = \partial_{\mu} A_{\nu} - \partial_{\nu} A_{\mu} + [A_{\mu},A_{\nu}]$. 

The action-extremizing Yang-Mills-Higgs field equations are easily derived, but we can more strongly constrain the solutions by requiring time-independence ($\partial_0=0$) and taking the Higgs mass to be vanishingly small (i.e. $\lambda \rightarrow 0$). Under these conditions, the energy is minimized when the following equation, called the Bogomolny equation, is satisfied:
\begin{equation}
B_i = \pm D_i \phi,
\end{equation}
where the magnetic field is found from the field strength: $B_i = - \frac{1}{2} \varepsilon_{ijk} F^{jk}$ and $i=1,2,3$. These conditions are collectively known as the \textit{BPS limit} and solutions to the Bogomolny equation are BPS magnetic monopoles \cite{Manton:2004tk}. In particular, we are interested here in exploring this theory in a three-space with two coordinates $x_1$ and $x_2$ compactified on a two torus, each with period $ 2\pi$: $(x_1,x_2) \sim (x_1 + 2\pi, x_2) \sim (x_1, x_2 + 2\pi)$, and $x_3 \in \mathbb{R}$. Monopoles in such a space are referred to as monopole walls, or \textit{monowalls}.

Certain components of the gauge field gain mass because the Higgs field is non-vanishing, and because of the gauge field holonomies associated with the periodic directions. As $x_3$ grows large, we choose the Higgs field to approach diagonal with at most linear growth, the gauge holonomies to approach diagonals which are constant in space, and the $U(N)$ symmetry to be maximally broken to $U(1)^N$ in the asymptotic region. Then only diagonal gauge field components, those representing the Cartan subalgebra of $U(N)$, remain massless. We identify the locations of magnetic charge with positions at which partial or full gauge symmetry is restored \cite{Fujimori:2008ee}. The massive gauge field components decay exponentially with distance from such charge. 

Now, a BPS solution is a static solution, i.e the Higgs and gauge field configurations are time-independent. For fixed total charge and a given set of gauge and Higgs field boundary conditions, there may be many such static solutions. A monopole (or monowall) moduli space is the set of BPS solutions for fixed total monowall charge and boundary conditions that together form a manifold. Each point on the manifold represents a BPS solution with associated charge distribution. If the positions of localized charge gain very small velocities, this motion can be approximated by geodesic motion on the moduli space. An additional effect comes with this small time-dependence: these magnetic charges gain electric charge and so altogether may interact magnetically, electrically, and via the scalar field. This effect is controlled by a periodic phase modulus $\theta$ associated with each charge \cite{Manton:1985hs}.

\subsection{Objectives }

This paper pursues two goals. The first goal is to show that a BPS monowall that has moduli (degrees of freedom) will split into distinct, well-separated sub-monowalls (or subwalls) if any of its moduli becomes large. The second goal is to determine the moduli space metric corresponding to the gauge field and Higgs interactions of $n$ well-separated, distinguishable, slow-moving sub-monowalls.

To accomplish the first objective, we will review the construction of the Higgs spectral curve and analyze its asymptotic behavior using the Newton polygon and amoeba associated with the curve. The amoeba asymptotics directly relate to the BPS monopole when its constituent charges are widely-spaced, so we will demonstrate that as one of the monowall's moduli becomes very large, the monowall breaks into subwalls which move apart. Furthermore, we show that the symmetry breaks from $U(N)$ to $U(1)^N$ at a determined distance from each subwall. The subwalls then behave as distinct charges and their gauge and Higgs field interactions are approximately abelian, with exponential precision.

We reach the second objective to calculate the moduli space metric for $n$ well-separated subwalls by modeling the moving subwalls as abelian planes with scalar, magnetic, and electric charge interacting with one another and with a set of background gauge and Higgs fields. For these subwalls the Lagrangian reduces to purely kinetic in the slow-move limit. Lagrange's equations produce the geodesic equation for the monowall moduli and we can read off the metric.

Here are the defining parameters of the moduli space we will calculate. The Yang-Mills-Higgs abelian asymptotic field equations imply a harmonic Higgs field. Following \cite{Cherkis:2012qs}, we constrain the Higgs field of the $U(N)$ monowall to diverge no more than linearly, and its eigenvalues to behave as follows when $x_3 \rightarrow \pm \infty$:
\begin{equation}
\phi_a^{\pm \infty} = -i \left( G^{\pm}_a x_3 + v^{\pm}_a \right) + \mathcal{O} ( x_3^{-1}) ,
\end{equation}
where $a = 1,...,N$ indexes the $N$ factors of $U(1)$, i.e. the $N$ diagonal elements of the field matrices with which the Higgs eigenvalues are in one-to-one correspondence. The left and right magnetic charges of the monowall $G^{\pm}_a $ are rational constants and the subleading terms $v^{\pm}_a$ are real constants. Also fixed as $x_3 \rightarrow \pm \infty$ are the holonomy eigenvalues $e^{i d_{1,a}}$ and $e^{i d_{2,a}}$ associated with the two periodic directions $(x_1,x_2)$. We use the shorthand $\vec{d}^{\pm}_a = (d^{\pm}_{1,a},d^{\pm}_{2,a},0)$, where the vector symbol indicates the three spatial directions and $d^{\pm}_{a,i} \in [0,2 \pi)$. Together with the locations of any singular (called Dirac) monowalls, these constants $(G^{\pm}_a,v^{\pm}_a, \vec{d}^{\pm}_a)$ fully specify the moduli space. Cherkis and Ward \cite{Cherkis:2012qs} have established consistency conditions which must be satisfied if BPS solutions are to exist. These are determined using the Newton polygon construction, which will be described later in this section. They determined\cite{Cherkis:2012qs} that the number of real moduli is then four times the number of integer points on the interior of the Newton polygon, which the next subsection describes.

\subsection{Higgs Spectral Curve}

For each periodic coordinate, define the \textit{Higgs spectral curve} (or ``monopole spectral curve'') \cite{Cherkis:2000cj, Cherkis:2000ft, Cherkis:2012qs}. We will use this tool to explore behaviors of BPS solutions. The $x_1$-direction Higgs curve $\Sigma_1$, for example, is determined by the characteristic equation for the holonomy of the differential operator $D_1 + i \phi$. The fields $(A_{\mu}, \phi)$ are assumed to be BPS. We will pursue the example of the $x_1$-direction Higgs curve but it should be noted that a different spectral curve could be found by simply exchanging the spatial indices $1$ and $2$. These curves share a Newton polygon, which we will shortly define and describe. To define the holonomy, introduce a matrix function $V(x_1,x_2,x_3)$ which solves the equation 
\begin{equation} \label{Holonomy equation}
(D_1 + i \phi) V = 0,
\end{equation}
with initial condition $V(0,x_2,x_3) = \mathbb{1}$. The holonomy of $(D_1 + i \phi)$ is $W(x_2,x_3) = V(2 \pi,x_2,x_3)$, which is a holomorphic function of $x_3 + ix_2$ \cite{Cherkis:2012qs}, given the $B_i = - D_i \phi$ form of the Bogomolny equation. Define a more convenient coordinate $s=e^{x_3 + ix_2}$. The eigenvalues of the holonomy $W(s)$ are finite and nonzero, the Higgs spectral curve is described by the characteristic (eigenvalue) equation of $W(s)$:
\begin{equation}
\begin{array}{ll}
\det(W(s) - t) = F(s,t) = 0, & \hspace{1.5cm} \text{where} \hspace{.3cm} F(s,t) = \sum\limits_l^N k_l(s) t^l .
\end{array}
\end{equation}
Given finite eigenvalues $t$ and the boundary conditions set on the fields in the previous section and in \cite{Cherkis:2012qs}, $F(s,t)$ is a polynomial in $t$ of degree $N$ and the functions $k_l(s)$ are rational functions of $s$. Without affecting the set of roots $\{(s,t) \}$ of $F$, we can rescale by a common denominator polynomial in $s$ to obtain a polynomial in $s$ and $t$, labeled $f(s,t)$. This is referred to as the spectral polynomial \cite{Cherkis:2012qs}, or \textit{Higgs spectral polynomial}. The curve produced by $f(s,t)=0$ is the \textit{Higgs spectral curve} and lives in $(\mathbb{C}^*)^2$, where $\mathbb{C}^*$ is the complex plane with the origin omitted, $s$ is the coordinate in the first factor of $\mathbb{C}^*$ and $t$ is the coordinate in the second factor.

\begin{figure}[htb]
\begin{center}
\centerline{\includegraphics[scale=.5]{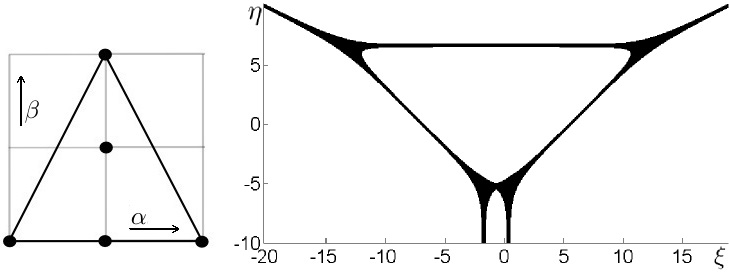}}
\caption{Newton polygon and amoeba for $ f(s,t)= 1.3 s t^2 + ust + 4 s^2 + 5 s - 1$ with $u=1000$.  \label{amoeba1}}
\end{center}
\end{figure}

We now introduce the Newton polygon and amoeba for this polynomial, which can be written $f(s,t)=\sum\limits_{i=0}^{\sigma} a_i s^{\alpha_i} t^{\beta_i}$, where $\sigma+1$ is the number of terms in the polynomial. The Newton polygon $\mathcal{N}(f)$ is the minimal convex hull of the points $ \{ (\alpha_i , \beta_i) \} $ in $\mathbb{Z}^2$ for which $a_i \neq 0$. The concept generalizes to arbitrary dimension \cite{Gaiotto:2009hg,Cherkis:2012qs}. To obtain the amoeba, project the Higgs spectral curve from two complex dimensions down to two real dimensions by taking the modulus of each factor of $\mathbb{C}^*$ and applying the Log map $(s,t) \rightarrow (\log|s|, \log|t|) = (x_3, \eta)$. This yields a more intuitive view of the nature of the curve, particularly in the large-$x_3$ regime, as will be seen. Asymptotically, Equation \eqref{Holonomy equation} simplifies significantly when the commutator vanishes and the Higgs field becomes approximately linear in $x_3$. It is clear that $x_3$ is the noncompact three-space coordinate and in this region  $\eta$ corresponds to the $x_3$-linear Higgs eigenvalue magnitudes. When the Higgs curve is projected in this manner, the result is called the \textit{amoeba} $\mathcal{A}(f) \in \mathbb{R}^2$ for its distinctive appearance \cite{Gelfand:1994tk} (see, for example, Figure $\eqref{amoeba1}$).

\section{Monowall Splitting} \label{3}

Each polygon edge is associated with a set of external amoeba legs stretching out to infinity. Each external leg is normal to its associated polygon edge, and its position is determined by the monomials of $f(s,t)$ associated with that edge (their powers $(\alpha,\beta)$ and coefficients). In order to keep the boundary conditions on the fields fixed, the polynomial coefficients corresponding to edge terms must be fixed \cite{Cherkis:2012qs}. This constraint does not apply to points on the interior of the Newton polygon, and we may consider the family of polynomials with fixed external coefficients and a range of values for internal coefficients. To this purpose, we begin by allowing one internal point coefficient to vary, i.e. we consider the family of polynomials for which the coefficient of one internal point takes any value on the complex plane except the origin, while the remaining coefficients are each fixed in the complex plane. Rather than considering each such polynomial individually, we may look at the whole picture at once by treating the internal coefficient as an independent variable on par with $s$ and $t$. This effectively increases the number of complex coordinates of the polynomial function from two to three. We will choose this varying internal coefficient $u = a_0$ to be associated with the lattice point $(\alpha_0,\beta_0)$ and write the three dimensional Higgs polynomial (from now on referred to as the Newton polynomial) with $\sigma+1$ terms as
\begin{equation}
\tilde{f}(s,t,u) = u \hspace{.1cm} s^{\alpha_0} t^{\beta_0} + \sum\limits_{i=1}^{\sigma} a_i s^{\alpha_i} t^{\beta_i}. 
\end{equation}

\begin{figure}[htb]
\begin{center}
\centerline{\includegraphics[scale=.5]{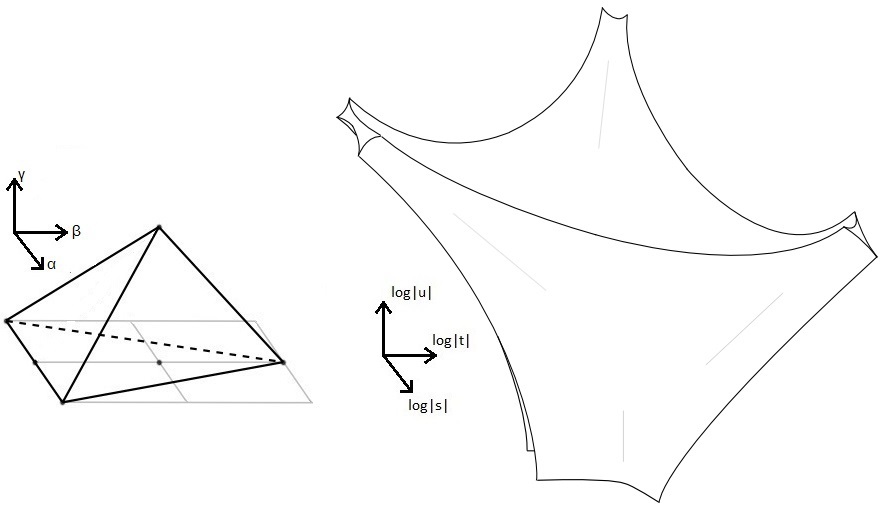}}
\caption{Three-dimensional Newton polygon and sketch of three-dimensional amoeba core for $ \tilde{f}(s,t,u) = 1.3 s t^2 + u s t + 4 s^2 + 5 s - 1$.  \label{amoeba2}}
\end{center}
\end{figure}

\begin{figure}[htb]
\begin{center}
\centerline{\includegraphics[scale=.5]{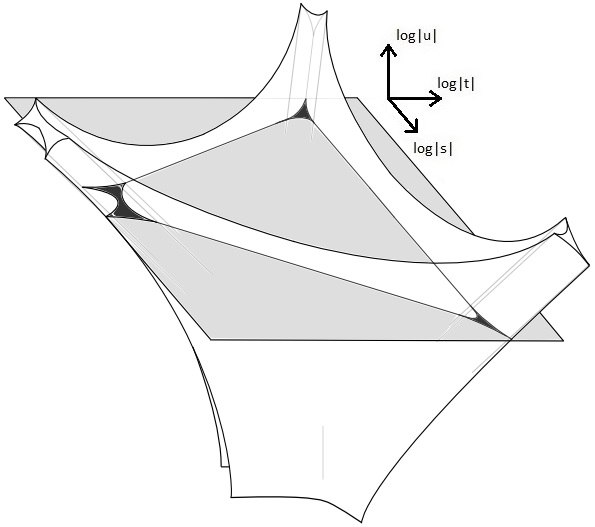}}
\caption{Three dimensional amoeba (white) and $R$ plane (grey). The intersection (black) gives the two-dimensional amoeba for a given value of $R$. \label{amoeba3}}
\end{center}
\end{figure}

The three-dimensional amoeba $\tilde{\mathcal{A}}(\tilde{f}) \in \mathbb{R}^3$ for $\tilde{f}(s,t,u) = 0$ also has externalities extending to infinity, known as the asymptotic three-dimensional amoeba. According to Gelfand, Kapranov, and Zelevinski\footnote{Proposition 1.13, Ch. 6} \cite{Gelfand:1994tk} and Viro \cite{Viro:2008xx}, this three-dimensional amoeba asymptotically exponentially approaches the \textit{core} of the amoeba, which can be described in the following way: Normal to each edge of the three-dimensional polytope for $\tilde{f}(s,t,u)$ are a continuous set of directions which form plane \textit{wedges}. Wedges for different edges on a face of the Newton polytope intersect at and and terminate on the \textit{leg} associated with that face. The three dimensional amoeba legs are a set of cylinders each normal to a polytope face and having two-dimensional amoeba cross-sections (see Figure $\eqref{amoeba2}$). Recall that $x_3 = \log|s|$ is the non-compact spatial coordinate and that $\eta = \log|t|$ asymptotically corresponds to the Higgs eigenvalue magnitudes. The new, third component $R = \log|u|$ is the non-compact modulus and its significance is seen in the intersection of the three-dimensional amoeba with a horizontal plane defined by a given height of $R$. This intersection is precisely the two-dimensional amoeba for $f(s,t)$ (e.g. Figure $\eqref{amoeba3}$). The Newton polygon for this two-dimensional amoeba is the projection of the three-dimensional Newton polytope onto the $(\alpha,\beta,0)$ lattice. Each subwall corresponds to a face of the three-dimensional polytope and corresponds to an edge of this two-dimensional polygon.

For a horizontal plane positioned at very large $R$, its intersection with the three-dimensional amoeba is as follows: The plane intersections with the wedges of the three-dimensional amoeba along straight lines, called \textit{amoeba lines}. Its intersections with the three-dimensional amoeba legs, called \textit{junctions}, are sections of two-dimensional-amoeba cylinders and, importantly, have fixed areas asymptotically which differ from the cylinder cross-sections by a constant factor. Each subwall, then, is asymptotically associated with and its behavior determined by a face of the Newton polytope. The separations/relative positions of subwalls \textit{depend linearly on the modulus $R$}. 

In the $\eta$ versus $x_3$ plane at $R$, the amoeba lines correspond to regions in $x_3$ where the Higgs eigenvalues take values linear in $x_3$, with degeneracy equal to the denominator of the slope. As we will show, the $U(N)$ symmetry in these regions is maximally broken to $U(1)^N$ by the non-vanishing gauge field holonomies for the $x_1$ and $x_2$ directions, and the fields are abelian. The junctions correspond to regions in which the Higgs field eigenvalues are not linear in $x_3$ and the gauge field holonomies cannot be approximated well, and so we are unable to infer fully broken symmetry; we interpret these regions as locations of magnetic charges, or subwalls. It is necessary now to define the widths of these subwalls, or the extents in $x_3$ of their nonabelian interiors. We will define the subwalls to be ``well-separated'' when their separations are much greater than the maximum subwall width and their interactions are abelian.

To accomplish this, we must quantify the decay of the non-commuting gauge field components which mediate nonabelian field interactions. Gauge field components which do not commute with the Higgs field must decay exponentially at a rate proportional to the separation of Higgs eigenvalues\footnote{\cite[Theorem 10.5, Ch. IV]{Jaffe:1980xx}}. Here this decay rate amounts to the Log of the ratio of eigenvalues, $\log \left( t_j/t_k \right)$, for the holonomy $\tilde{W}(s,u)$ since nonvanishing gauge field holonomies can asymptotically generate gauge field masses analogously to the Higgs mechanism. At the point where these non-commuting gauge field components have decayed by some chosen fraction, we mark the edge of a subwall. We define the subwall width as the distance at which the exponential rates for the decay of the nonabelian gauge field components are bounded from below by some small value $T_0$, plus the distance $1/T_0$ at which the fields will have decreased by a factor of $1/e$.

While the Higgs eigenvalue behavior (as a function of $x_3$) is illustrated by the amoeba, the behavior of the gauge field holonomy is not. We must therefore look to the spectral polynomial to determine the various branches of $t=T(s,u)$, which locally satisfy $\tilde{f}(s,T(s,u),u)=0$. This is done by calculating the Newton-Puiseux expansion \cite{McDonald, Beringer} for $T(s,u)$ with respect to $s$ and $u$. If the Newton polytope faces corresponding to two subwalls are adjacent, then the fields between two subwalls are governed primarily by the two monomials in the spectral polynomial that are associated with the edge $e$ joining the two faces. There are also smaller contributions from the remaining monomials. The resulting expansion will take the following form and only the first two terms in the expansion are of concern here:
\begin{equation}
T_j(s,u) = c_{1j} \hspace{.1cm} s^{\gamma_1} u^{\gamma_3} + c_{2j} \hspace{.1cm} s^{\tilde{\gamma}_{1j}} u^{\tilde{\gamma}_{3j}} + ... = c_{1j} \hspace{.1cm} s^{\gamma_1} u^{\gamma_3} \left( 1 + \left( c_{2j} / c_{1j} \right) s^{\tilde{\gamma}_{1j}-\gamma_1} u^{\tilde{\gamma}_{3j} - \gamma_3} \right) + ...
\end{equation}
\begin{figure}[htb]
\begin{center}
\centerline{\includegraphics[scale=.35]{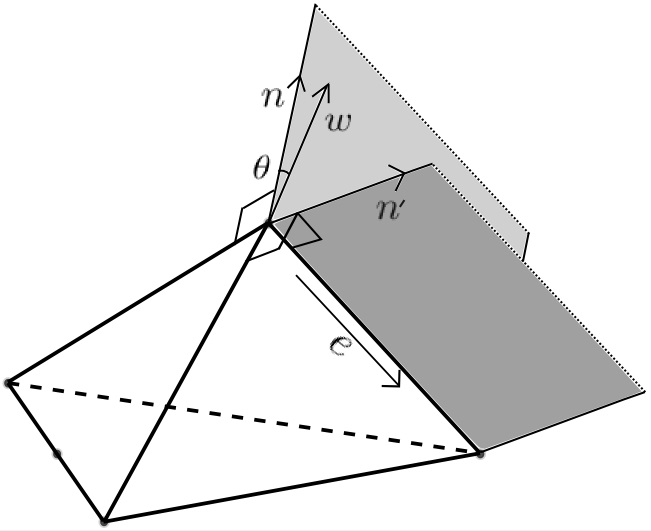}}
\caption{ \label{Amoeba4} The area above the grey partial planes is the normal cone for edge $e$. The normal vector $n'$ is normal to the front right face, while $n$ is normal to the rear face. The vector $w$ is normal to edge $e$ and lies in the wedge bounded by $n$ and $n'$. It is defined as a rotation if $n$ through angle $\theta$. \label{Amoeba4}}
\end{center}
\end{figure}
Briefly, for a direction $w \in \mathbb{R}^3$ within the normal cone of an edge of the Newton polytope (see Figure \ref{Amoeba4}), the Newton-Puiseux series is constructed iteratively. The first series term solves the vanishing of the edge polynomial $u s^{\alpha_0} T^{\beta_0} + a \hspace{.05cm} s^{\alpha} T^{\beta}=0$, so that in the first term in the series, the coefficient is $c_{1j} = (-a)^{-1/(\beta - \beta_0)} e^{2 \pi i \cdot j/ (\beta - \beta_0)}$, and the powers are $\gamma_1 = - (\alpha - \alpha_0) / (\beta - \beta_0)$, and $\gamma_3 = 1/(\beta - \beta_0)$. More formally, the powers $\gamma_1$ and $\gamma_3$ are the negative of the components of the slope vector $S_e = \left( \frac{e_1}{e_2}, 0, \frac{e_3}{e_2} \right) = -(\gamma_1, 0, \gamma_3)$ associated with the edge $e = (\alpha - \alpha_0, \beta - \beta_0, -1)$, and the coefficient $c_1$ solves the equation $\sum\limits_{i \in e} a_i c_1^{\beta_i}=0$, excluding trivial solutions. The second term, $c_2 s^{\tilde{\gamma}_1} u^{\tilde{\gamma}_3}$ is found by repeating this process for the Newton polytope for the polynomial $\tilde{f}_1(s,T_1,u) = \tilde{f}(s,T_1+c_1 s^{\gamma_1}u^{\gamma_3}, u) = \sum\limits_{i=0}^{\sigma^1} a^1_i s^{\alpha^1_i} T_1^{\beta^1_i}$, choosing an edge $\tilde{e}$ which maximizes $-S_{\tilde{e}} \cdot w$ (called the \textit{order} $\tilde{\nu}$ of edge $\tilde{e}$ with respect to $w$) while satisfying $-S_{\tilde{e}} \cdot w < - S_e \cdot w $. The coefficient $c_2$ solves the equation $\sum\limits_{i \in \tilde{e}} a^1_i c_2^{\beta^1_i}=0$ and $|c_2| \leq \left(1 + \frac{\text{max}\left( \{ |a_i c_1^N| \}, \{ |a_i| \} \right)}{\text{min}\left( \{ |a_i c_1^N| \} , \{ |a_i| \} \right)} \right) = : C_2$ is its maximum magnitude \cite{Estermann}.

Define $\delta = (\delta_{1j},0,\delta_{3j})=(\tilde{\gamma}_{1j} - \gamma_1,0,\tilde{\gamma}_{3j} - \gamma_3)$, which behave as follows in the asymptotic limits: For $s \rightarrow 0$ and $u \rightarrow \infty$, $\delta_{1j }> 0$ and $\delta_{3j} <0$; for $s \rightarrow \infty$ and $u \rightarrow \infty$, $\delta_{1j} < 0$ and $\delta_{3j} <0$. In other words, the quantity $s^{\delta_{1j}} u^{\delta_{3j}}$ decays in both of these limits of $s$ and $u$. Given the first two terms of the Newton-Puiseux series, the ratio of two eigenvalues $T_j$ of the holonomy $\tilde{W}(s,u)$ is written
\begin{equation} \label{8}
\frac{T_{j}(s,u)}{T_{k}(s,u)} = \frac{c_{1j}}{c_{1k}} \left( 1 + \frac{c_{2j}}{c_{1j}} s^{\delta_{1j}} u^{\delta_{3j}} - \frac{c_{2k}}{c_{1k}} s^{\delta_{1k}} u^{\delta_{3k}} \right) + \mathcal{O} \left( \text{min}( s^{2\delta_{1i}} u^{2\delta_{3i}}) \right)_{i=j,k}
\end{equation}
In this expression, every quantity but the first term $c_{1j}/c_{1k}$ decays in the asymptotic limits. Simplifying the ratio of coefficients $c_{1j}/c_{1k} = e^{2 \pi i (j-k)/(\beta - \beta_0)}$, the Log of equation \eqref{8} becomes 
\begin{equation} \label{9}
\log \left( \frac{T_{j}}{T_{k}} \right)(s,u) = \frac{2 \pi i (j-k)}{\beta - \beta_0} + \left( \frac{c_{2j}}{c_{1j}} s^{\delta_{1j}} u^{\delta_{3j}} - \frac{c_{2k}}{c_{1k}} s^{\delta_{1k}} u^{\delta_{3k}} \right) +\mathcal{O} \left( \text{min}( s^{2\delta_{1i}} u^{2\delta_{3i}}) \right)_{i=j,k}
\end{equation}
The first term in this series is constant, while in the asymptotic limit the quantity in the curved brackets is the largest decaying term in the series.

The expansion direction $w \in \mathbb{R}^3$ comes explicitly into play when determining the relative sizes of the quantities $s^{\delta_1}$ and $u^{\delta_3}$. Along the direction $w$, the variables behave as $(s_0,t_0,u_0) \rightarrow (s_0 e^{\bar{w}_1},t_0 e^{\bar{w}_2}, e^{\bar{w}_3}) $ relative to some initial values $(s_0,t_0,1)$ \cite{Viro:2008xx}, where $\bar{w}$ is the vector $w$ multiplied by a coefficient so that its third component is $\bar{w}_3=R$: $\bar{w} = \frac{R}{w_3} w$ for $R \in \mathbb{R}^+$. Also define the extended face normal vector $\bar{n} = \frac{R}{n_3} n$. We have not said very much so far about the direction vector $w$ except that it must lie within the normal cone of the edge $e$. Define it in terms of the normal vector $n$ for one of the edge's adjacent faces (see Figure \eqref{Amoeba4}). For angle $\theta$, we define the expansion direction as a rotation of the nearest of the two adjacent face normal vectors: $w=n \cos \theta + \frac{(e \times n)}{|e|} \sin \theta + \frac{e(e \cdot n)}{|e|^2}(1-\cos \theta)$. The third term vanishes since the face normal $n$ is orthogonal to the edge vector and $e \cdot n =0$. Applying this form for the vector $w$, the largest decaying terms in Equation \eqref{9} are
\begin{equation}
s^{\delta_{1j}} u^{\delta_{3j}} = s_0^{\delta_{1j}} e^{(\tilde{\nu} - \nu) R/w_3} = s_0^{\delta_{1j}} e^{-|\bar{w}_1 - \bar{n}_1|/\lambda},
\end{equation}
where the denominator $\lambda := \frac{|(e \times n)_1 - (e \times n)_3 n_1/n_3 |}{|(e \times n) \cdot \delta|}$ is $j$-independent, and the powers are $\delta = (\delta_{1j}, 0 , \delta_{3j})$ for any $j$. The difference in orders of the secondary edge $\tilde{e}$ and the original edge $e$ is $(\tilde{\nu} - \nu) = w \cdot \delta$ and it is $j$-independent. The vector component $\bar{n}_1 = x_{3l} - x_{3l}^0$ is the $x_3$-distance between the subwall's position $x_{3l}$ and its reference position $x_{3l}^0$, i.e. the linearly extrapolated position of the wall when $R=0$. We identify the subwall initial position for edge $e$ with the greatest magnitude as $\text{max}|x_{3,l}^0|$ and that with the smallest magnitude as $\text{min}|x_{3l}^0|$ for $l = 1,...,n$. 

For a $U(N)$ monowall, we find that beyond a distance $\lambda \log\left|\frac{c_{jk} e^{N/ \lambda \pi}}{\pi/N} \right|$ from the wall's position, the exponential decay rates of the off-diagonal gauge field components are bounded by $\left| \log(T_{j}/T_{k}) \right| \geq \pi / N$, where the mixed-index coefficient is defined $c_{jk} := \left( \frac{c_{2j}}{c_{1j}}s_0^{\delta_{1j}} - \frac{c_{2k}}{c_{1k}}s_0^{\delta_{1k}} \right)$ and the power of $s_{0l}$ is bounded by $1/N^2 \leq |\delta_{1j}| \leq N^2$. The bounded Newton-Puiseux coefficients satisfy $|c_2| \leq \left( 1 + \frac{\text{max} \left( \{ |a_i c_1^N| \},\{|a_i| \} \right) }{\text{min} \left( \{ |a_i c_1^N| \},\{|a_i| \} \right) } \right) = : \mathbb{C}_2$ and $|c_1| \geq \text{max}\{ |a_i | \}^{-\rho} = : \mathbb{C}_1$, where $\rho = N$ if $\text{max}\{ |a_i| \}$ is greater than unity and $\rho=1/N$ if it is less than unity. The coefficient $c_{jk}$ is then limited by $|c_{jk}| \leq 2 \frac{\mathbb{C}_2}{\mathbb{C}_1} e^{x_3^0 \delta_1}$. We find the following is the maximum subwall width for a $U(N)$ monopole on $\mathbb{R} \times T^2$ where each torus period is $2 \pi$:
\begin{equation}
\frac{\ell}{2} = \log \left| \left( \frac{2 N \mathbb{C}_2 }{\pi \mathbb{C}_1} \right)^{\lambda} \right| + \text{max}|x_{3j}| N^2 \lambda + \frac{N}{\pi} + \left( \text{max}(x_{3j}^0) - \text{min}(x_{3j}^0) \right),
\end{equation}
where $\lambda := \frac{|(e \times n)_1 - (e \times n)_3 n_1/n_3 |}{|(e \times n) \cdot \delta|}$. No subwall of a $U(N)$ monowall may have a width greater than $\ell$. Recall that the subwall separations are linear in $R$. In order to consider the subwalls to be well-separated and their interactions abelian, we require of the non-compact modulus
\begin{equation}
R >> \ell
\end{equation} 
for each $U(N)$ subwall.

In this section, we described the asymptotic behavior of a BPS $U(N)$ monowall using the Higgs spectral curve and its Newton polygon and amoeba. By allowing a coefficient in the interior of the polygon to vary, we introduced a modulus $R = \log|u|$ for the monowall. Using the Newton-Puiseux series for the eigenvalues $T_j(s,u)$ of the holonomy $\tilde{W}(s,u)$ to characterize the off-diagonal gauge field decay rates, we showed that when the modulus $R$ is very large, the monowall breaks into subwalls whose separations increase with $R$. For $R >> \ell$, the subwall interactions reduce to $N$ abelian interactions (i.e. $U(N)$ breaks to $U(1)^N$) up to corrections exponentially small in $R$. In the following section, we will derive an asymptotic moduli space metric for these well-separated subwalls using their abelian scalar, magnetic, and electric interactions.

\section{Asymptotic Moduli Space} \label{4}

\subsection{Lagrangian and Fields}

We have established that a monowall splits into subwalls when a modulus becomes large. We will now consider the regime in which the monowall is split into $n$ subwalls which have no internal moduli, and the sub-wall Higgs and gauge interactions are abelian. In order to model these abelian long-distance interactions, we contrive a system of $n$ abelian monowalls in a linear Higgs background which have scalar, magnetic, and electric interactions with one another and with the background. The interactions of one of these abelian monowalls with the background and the $n-1$ remaining abelian monowalls mimics the long-distance interactions of one of the nonabelian subwalls with the $n-1$ remaining subwalls. We will from here forward refer to these model abelian monowalls simply as subwalls. We describe the abelian monowall interactions using Lorentz-invariant Maxwell electromagnetism with a scalar field. We write the Lagrangian and consider only very small subwall velocities. We then Legendre transform from the electric charge $q_i$, which is a momentum, to its canonical coordinate, which is a periodic phase modulus $\theta_i$. The Lagrangian reduces to purely kinetic under these conditions. We will then read the monowall moduli space metric off of this kinetic Lagrangian. 

Recall that we choose a gauge in which the Higgs field is diagonal and we have demonstrated that it is $x_3$-linear for large $x_3$. The off-diagonal gauge fields gain mass and are exponentially small and therefore negligible, while the diagonal gauge field components remain massless. We will represent the generators of the Cartan subalgebra as the $N$ generators $\{ H_a \}$ of $N \times N$ imaginary diagonal matrices, and write the asymptotic fields as $\phi = \phi^a H_a$ and $A_{\mu} = A^a_{\mu} H_a$ where $(\phi^a,A_{\mu}^a)$ are real and $a=1,...,N$. We employ an additional, adjoint dual gauge potential $\tilde{A}_{\mu} = \tilde{A}^a_{\mu} H_a$ to model magnetic interactions. This dual gauge field can be related to $\vec{A}_{\mu}$ via the usual dual field strength $\tilde{F}_{\mu \nu} = \partial_{\mu} \tilde{A}_{\nu} - \partial_{\nu} \tilde{A}_{\mu} + [\tilde{A}_{\mu}, \tilde{A}_{\nu}]$, which is defined as $\tilde{F}_{\mu \nu} = - \frac{1}{2} \varepsilon_{\mu \nu \rho \sigma} F^{\rho \sigma}$. The relativistic Lagrangian for the $i^{th}$ wall interacting with the gauge, dual gauge, and Higgs fields $(\phi,A_{\mu}, \tilde{A}_{\mu})$ of the $n-1$ remaining subwalls and background is
\begin{equation}
L_i = -i\text{Tr} \left[ 4 \pi \phi \sqrt{g_i^2 + q_i^2} \sqrt{1 - \vec{V}_i^2} - 4 \pi q_i A^0 + 4 \pi q_i \vec{V}_i \cdot \vec{A} - 4 \pi g_i \tilde{A}^0 + 4 \pi g_i \vec{V}_i \cdot \vec{\tilde{A}} \right],
\end{equation}
where the three-space velocity is $\vec{V}_i = \dot{\vec{x}}_i$ and we use the dotted time-derivative notation $\dot{x} = \frac{d x}{dt}$. The magnetic, electric, and scalar charges of the $i^{th}$ subwall are interpreted as $(g_i, q_i, \sqrt{g_i^2 + q_i^2})$ respectively, where this form of the scalar charge follows from the BPS conditions under which the static forces cancel for well-separated subwalls \cite{Gibbons:1995yw}. Note that the electric charges $q_i$ are momenta associated with the phase degrees of freedom $\theta_i$ for subwalls. The electric charges of the subwalls are subject to net electric charge conservation, and individual electric charge conservation when the subwalls are well-separated as they are here.

Recall that we allow gauge field holonomies to have non-zero, spatially uniform, linearly time-dependent values asymptotically. Define the linearly time-dependent terms in the asymptotic holonomies (phases) $\vec{a}(t)=\text{sgn}(x_3)(a,b,0)(t)$ associated with each of the periodic spatial directions, and their dual vector $\vec{\tilde{a}}(x)$ such that $\dot{\vec{a}} = \vec{\nabla} \times \vec{\tilde{a}}$. The effect of the phase velocities $\dot{\vec{a}}$ is equivalent to that of the transverse spatial velocities $\dot{x}_{1i}$ and $\dot{x}_{2i}$: $\vec{V}_i = (- \dot{b}_i, \dot{a}_i, \dot{x}_{3i})$. For later use, define two dual functions $u(x_3)$ and $\vec{w}(x)$ such that $\vec{\nabla} u = \vec{\nabla} \times \vec{w}$. 
\begin{equation}
\begin{array}{ll}
\vec{w}(x) = \vec{w}(-x) = \frac{1}{2} \text{sgn}(x_3) \left[ -x_2 \hat{x}_1 + x_1 \hat{x}_2 \right], & \hspace{.8cm} \vec{a}(t) =  \text{sgn}(x_3)( a(t) \hat{x}_1 + b(t) \hat{x}_2  ) , \\
u(x_3) = u(-x_3) = |x_3|, &  \hspace{.8cm} \vec{\tilde{a}}(x) = \text{sgn}(x_3) [ - \dot{b} x_1 + \dot{a} x_2 ] \hat{x}_3.
\end{array}
\end{equation}

In addition to fields generated by subwalls, which we will write next, we include static field backgrounds for each factor of $U(1)$. For convenience, we split these backgrounds into constant terms $(d_{\mu}^a, \tilde{d}_{\mu}^a,-v^a)$ and a background linear Higgs $ \phi_0$ with the associated linear gauge fields $(A_{\mu,0},\tilde{A}_{\mu,0})$ required by Bogomolny's equation:
\begin{equation}
\begin{array}{lll}
(d_{\mu}, \tilde{d}_{\mu},-v) = (d_{\mu}^a H_a, \tilde{d}_{\mu}^a H_a,-v^a H_a), & \hspace{.5cm} A_0^{0,a}(x) = 0,  & \hspace{.5cm} \phi_0^a(x_3) = -g_0^a x_3, \\
\vec{A}_0^a(x) = \frac{g_0^a}{2} \left[ -x_2 \hat{x}_1 + x_1 \hat{x}_2 \right], 
& \hspace{.5cm} \vec{\tilde{A}}_0^a(x) = 0, 
& \hspace{.5cm} \tilde{A}^{0,a}_0(x) = - g_0^a x_3.
\end{array}
\end{equation}

The gauge and Higgs fields for the $j^{th}$ wall moving with velocity $\vec{V}_j$ are Lorentz boosted versions of those for the stationary wall. We keep only terms up to quadratic in velocities and electric charge in the Lagrangian, so in the fields $\vec{A}$ and $\vec{\tilde{A}}$ discard terms which are higher order than linear in velocities. Similarly, in the scalar expressions $\phi$, $A^0$, $\tilde{A}^0$ discard terms which are higher order than quadratic in velocities. This requires approximation of the Li\'enard-Wiechert denominator $ (\vec{x}^2 - ( \vec{x} \times \vec{V} )^2 )^{1/2}$ as $|\vec{x}|$ since the denominator would appear in the scalar-type quantities with coefficients linear in velocity, resulting in negligible terms cubic in velocity \cite{Cherkis:2001gm, Gibbons:1995yw}.
In this approximation, a subwall moving at velocity $\vec{V}_j$ with respect to the origin generates the following fields $(\phi_j^a, A_{\mu,j}^a, \tilde{A}_{\mu,j}^a)$. 
\begin{align}
& \phi_j^a(x_3) = -  u(x_3) \sqrt{ (g_j^a)^2 + q_j^2 } \sqrt{ 1 - \vec{V}_j^2 } , \nonumber \\[.2cm]
& A^{0,a}_j(x) =  - q_j u(x_3) + g_j^a \left(\vec{w}(x) \cdot \vec{V}_j \right) - g_j^a \left( \vec{a} \cdot \vec{V}_j \right) , \nonumber \\[.2cm]
& \tilde{A}^{0,a}_j(x) = - g_j^a u(x_3) - q_j \left( \vec{w}(x) \cdot \vec{V}_j \right) - g_j^a \left( \vec{\tilde{a}}_j(x) \cdot \vec{V}_j \right) , \\[.2cm]
& \vec{A}_j^a(x) = g_j^a \vec{w}(x) - g_j^a \vec{a}, \nonumber \\[.2cm]
& \vec{\tilde{A}}_j^a(x) =  - q_j \vec{w}(x) - g_j^a \vec{\tilde{a}}_j(x) - g_j^a u(x_3) \vec{V}_j . \nonumber
\end{align}

The net gauge fields must respect the periodic boundary conditions on $\mathbb{R} \times T^2$ and so we require that, for a coordinate shift in one of the periodic directions, the fields be gauge-shifted under the $U(1)$ symmetry, with gauge functions given here. 
\begin{equation} \label{BCs}
\begin{array}{l}
x_1 \rightarrow x_1+2 \pi, \hspace{.9cm} \theta_j \rightarrow \theta_j + \pi g_j \text{sgn}(x_3)x_2, \\
x_2 \rightarrow x_2+2 \pi, \hspace{.9cm} \theta_j \rightarrow \theta_j - \pi g_j \text{sgn}(x_3) x_1.
\end{array} 
\end{equation}

\subsection{Two-Monowall Interactions}

Using these fields, we may now write the Lagrangian in a convenient form and begin by doing so for two subwalls. For a pair of walls, define the following relative position, phase, and charge quantities:
\begin{equation}
\begin{array}{lll}
\vec{x} = \vec{x}_1-\vec{x}_2, & \hspace{.8cm} q^a = \frac{q_1}{g^a_1} - \frac{q_2}{g^a_2}, & \hspace{.8cm} g^a = g_1^a - g_2^a, \\
\vec{a} = \vec{a}_1 - \vec{a}_2, & \hspace{.8cm} \vec{\tilde{a}} = \vec{\tilde{a}}_1 - \vec{\tilde{a}}_2,  & \hspace{.8cm} G^a = \sum\limits_{i=1}^2 g_i^a.
\end{array}
\end{equation}
Neglecting constant terms, suppressing the index $a$ for $U(1)$ factors, the symmetrized Lagrangian for each set of $U(1)$ interactions takes the form
\begin{align}
\frac{L}{4 \pi} = & \frac{v}{2G} \left[ \left( \sum\limits_{i=1}^2 g_i \vec{V}_i \right)^2  +  g_1 g_2 \vec{V}^2    \right] -  
 \frac{v}{2G} \left[ \left( \sum\limits_{i=1}^2 g_i \frac{q_i}{g_i} \right)^2  +  g_1 g_2 q^2  \right] + \sum\limits_{i=1}^2 q_i \vec{V}_i \cdot \vec{d} - \\ 
& - \sum\limits_{i=1}^2 q_i d^0
+  \sum\limits_{i=1}^2 g_i \vec{V}_i \cdot \vec{\tilde{d}} + \sum\limits_{i=1}^2 \frac{g_0 g_i x_{3,i}}{2} \left( \vec{V}_i^2 - \frac{q_i^2}{g_i^2} \right) + \sum\limits_{i=1}^2 g_0 g_i \frac{q_i}{g_i} \left(\vec{w}(x) \cdot \vec{V}_i \right) \nonumber\\
& + \left[  \frac{g_1 g_2 u(x_3)}{2} (\vec{V}^2 - q^2) + g_1 g_2 q \left( \vec{w}(x) \cdot \vec{V} \right)  + g_1 g_2 [\vec{\tilde{a}} - q \vec{a} ] \cdot \vec{V} \right] . \nonumber
\end{align}
To find the full Lagrangian, we add up all $N$ of these $U(1)$ Lagrangians. This splits into the center of mass Lagrangian and the remainder Lagrangian, $L = L_{CM} + L_{rem}$. We integrate here over the periodic coordinates $x_1$ and $x_2$ from $-\pi$ to $\pi$. Because the terms with $\vec{w}(x) \cdot \vec{V}$ and $\vec{\tilde{a}}(x) \cdot \vec{V}$ are linear in $x_1$ and $x_2$ positions, these terms vanish after integration.\footnote{Altering the integration bounds of $x_1$ and $x_2$ yields different but physically equivalent forms of the Lagrangian.} Here is the result after separating the center of mass and remainder components of the Lagrangian, with implicit sum over the suppressed index $a$:
\begin{align}
\frac{L_{CM}}{4 \pi} = & \frac{v}{2G} \left[ \left( \sum\limits_{i=1}^2 g_i \vec{V}_i \right)^2 - \left( \sum\limits_{i=1}^2 g_i \frac{q_i}{g_i} \right)^2 \right] 
+ \frac{1}{2} \left( \sum\limits_{i=1}^2 \frac{q_i}{g_i} \right) \left( \sum\limits_{j=1}^2 g_j \vec{V}_j \cdot \vec{d} \right)  - \nonumber \\
& - \left( \sum\limits_{i=1}^2 g_i \frac{q_i}{g_i} d^0 \right)  + \left( \sum\limits_{i=1}^2 g_i \vec{V}_i \cdot  \vec{\tilde{d}}  \right), \\[.5cm]
\frac{L_{rem}}{4 \pi} = & \frac{g_1 g_2}{2} \left( \frac{v}{G} + u(x_3) \right) (\vec{V}^2 - q^2) + \sum\limits_{i=1}^2 \frac{g_0 g_i x_{3,i}}{2} \left( \vec{V}_i^2 - \frac{q_i^2}{g_i^2} \right) -  \nonumber\\
& - g_1 g_2  ( \vec{a} \cdot \vec{V} ) q + \frac{\vec{d}}{2} \cdot \left( g_1 \vec{V}_1 - g_2 \vec{V}_2 \right) q. \nonumber
\end{align}

Maintaining the low-velocity approximation, the Lagrangian is purely kinetic since $q$ behaves as a velocity. We now apply the fixed asymptotic boundary conditions constraint, which is equivalent to fixing the sums of the three-space and periodic positions of the subwalls (i.e. fixing the center of mass, or its analog). Incidentally, there is a physical motivation for fixing the boundary conditions. Because the fields diverge as $x_3 \rightarrow \pm \infty$, so too does the energy. Changing the boundary conditions on the fields would require infinite kinetic energy. After fixing the center of mass, the Lagrangian reduces to the remainder Lagrangian (from now on referred to simply as the Lagrangian).

Currently, the Lagrangian is written in terms of the $x_3$ positions of the subwalls, their phases $\vec{a}$, and their electric charges. The electric charge is not itself a modulus but is a momentum conjugate to a periodic modulus $\theta$. A Legendre transform, changing coordinates from $q$ to $\dot{\theta}$, produces the Lagrangian written explicitly in terms of the monowall moduli.
\begin{equation}
L' = L + q \dot{\theta}.
\end{equation}
After implementing the Legendre transform, we write the metric in Lee-Weinberg-Yi form \cite{Lee:1996kz} in terms of absolute rather than relative coordinates.
\begin{equation} \label{TheMetric}
\boxed{ \frac{ds^2}{4 \pi} = \sum\limits_{a=1}^N \hspace{.1cm}\frac{1}{2} U_{ij,a} d\vec{x}^i \cdot d\vec{x}^j + \frac{1}{2} \left[ U^{-1} \right]_{ij,a} 
 \left[ d\theta^i + \vec{W}^{ik}_a \cdot d\vec{x}_k \right] 
 \left[ d\theta^j + \vec{W}^{jl}_a \cdot d\vec{x}_l 
 \right]
 }
\end{equation}
with the following tensors defined for two subwalls:
\begin{equation}
\begin{array}{ll}
U_{ii,a} = g^0_a g_{ia} x_{3i} + \sum\limits^2_{\substack{ j = 1 \\ j \neq i}} g_{ia} g_{ja} \left( \frac{v_a}{G_a} + |x_{3i} - x_{3j} | \right), & \vec{W}^{ii}_a = \frac{\vec{d}_a}{2} g^i_a - \sum\limits^2_{\substack{ j = 1 \\ j \neq i}} g^i_a g^j_a (\vec{a}^i - \vec{a}^j ), \\
U_{ij,a} = - g_{ia} g_{ja}  \left( \frac{v_a}{G_a} + |x_{3i} - x_{3j}| \right), & \vec{W}^{ij}_a = -\frac{\vec{d}_a}{2} g^j_a + g^i_a g^j_a  (\vec{a}^i - \vec{a}^j ).
\end{array}
\end{equation}
where the third components of the following vectors vanish $d_{3a}=W^{ii}_{3a} = W^{ij}_{3a}=0$ and the three-space differential is expressed $d\vec{x} = (- db, da, dx_3)$. The index $a=1,...,N$ runs over the factors of $U(1)$. This metric retains the $U(1)$ symmetries, and symmetry under the $SL(2,\mathbb{Z})$ action on the $x_1$ and $x_2$ phases $(a,b)$.

\subsection{Multi-Monowall Interactions and Moduli Relations}

The metric $\eqref{TheMetric}$ holds for the extension to $n$ subwalls. The $n$-subwall tensors are
\begin{equation}
\begin{array}{ll}
U_{ii,a} =  g^0_a g_{ia} x_{3i} + \sum\limits^n_{\substack{ j = 1 \\ j \neq i}} g_{ia} g_{ja} \left( \frac{v_a}{G_a} + |x_{3i} - x_{3j} |  \right), &  \vec{W}^{ii}_a = \frac{\vec{d}_a}{n} g^i_a - \sum\limits^n_{\substack{ j = 1 \\ j \neq i}} g^i_a g^j_a (\vec{a}^i - \vec{a}^j ), \\
U_{ij,a} = - g_{ia} g_{ja}  \left( \frac{v_a}{G_a} + |x_{3i} - x_{3j} |  \right), & \vec{W}^{ij}_a = -\frac{\vec{d}_a}{n} g^j_a + g^i_a g^j_a  (\vec{a}^i - \vec{a}^j ).
\end{array}
\end{equation}
where $d^3_a=W^{ii}_{3,a} = W^{ij}_{3,a}=0$ and $d\vec{x} = (-db, da, dx_3)$. The index $a=1,...,N$ again runs over the factors of $U(1)$.

When the consistency conditions of \cite{Cherkis:2012qs} are applied, the number of independent moduli reduces from $4n$ to $4 \Gamma$, where $\Gamma$ is the number of internal points in the Newton polygon. To illustrate this for the above BPS monowall moduli space, we first consider the case of one varying internal coefficient. We use the modulus $R$ to parameterize the values of $x_3$ and $\phi$ in the $R$-plane corresponding to each of the $n$ amoeba junctions. This is done by finding the lines in $\mathbb{R}^3$ along which two adjacent three-dimensional amoeba wedges intersect and using these to define subwall positions for each value of $R$. We will from here forward refer to the two-dimensional amoeba, which is the amoeba for the projection of the three-dimensional Newton polytope onto the $(\alpha,\beta,0)$ lattice. For each value of $R$, there is a different two-dimensional amoeba. Recall that each subwall corresponds to a face of the three-dimensional polytope and therefore each subwall now corresponds to an edge of this two-dimensional polygon. The relationships between $R$ and $(x_3,\phi)$ at the junctions are linear:
\begin{equation}
\begin{array}{ll}
dx_3^i = m^i \hspace{.1cm} dR  , & \hspace{1.5cm} d\phi^i = n^i \hspace{.1cm} dR.
\end{array}
\end{equation}
The coefficients $(m^i,n^i)$ can be found by direct examination of the amoeba since $n_i/m_i$ is the slope of the $i^{th}$ external amoeba leg. They can be expressed in terms of the perimeter points of the Newton polygon. Let the lattice coordinates of the $i^{th}$ vertex in the Newton polygon be written $(\alpha^i,\beta^i) \in \mathbb{Z}^2$, with $i \in \mathbb{Z}/n$ running counterclockwise over the $n$ vertices of the Newton polygon. Asymptotically, each sub-triangle in the Newton polygon triangulation represents a subwall and, as we will show, can be used to determine its motion. We choose a triangulation such that each sub-triangle contains an internal point $(\alpha_0,\beta_0)$, and label the remaining two vertices $(\alpha^i,\beta^i)$ and $(\alpha^{i+1},\beta^{i+1})$ for the $i^{th}$ sub-triangle. 

Under the Log map, the set of solutions to the polynomials for each of the sub-triangle edges containing $(\alpha_0,\beta_0)$ form wedges which intersect along a line. This line represents the set of positions the associated subwall may occupy in the $x_3-\eta$ plane for all values of the modulus $R \in \mathbb{R}^+$. The set of common solutions can be found by simultaneously solving the polynomials for two of its edges. In the following linear equation, the first row corresponds to the sub-triangle edge connecting $(\alpha_0,\beta_0)$ to $(\alpha_i,\beta_i)$, and the second row corresponds to $(\alpha_0,\beta_0)$ and $(\alpha_{i+1},\beta_{i+1})$. We define a reference point $(x_3^0,\eta^0)$ by solving this equation for the case $R=0$. Then, we can relate $x_3^i$ and $\eta^i$ to each modulus $R$ in the following way:
\begin{equation}
\begin{array}{ll}
\small dR \left( \begin{matrix} 1 \\ 1 \end{matrix} \right) = \left( \begin{matrix} \alpha_i - \alpha_0 & \beta_i - \beta_0 \\ \alpha_{i+1} - \alpha_0 & \beta_{i+1} - \beta_0 \end{matrix} \right) \left( \begin{matrix} dx_3 \\ 2 \pi d\phi  \end{matrix} \right)_i, \normalsize & \hspace{1cm}
\begin{array}{l}
dx_3^i= \frac{\beta_{i+1} - \beta_i}{\det^i} dR, \\[.2cm] d\phi^i = \frac{-(\alpha_{i+1} - \alpha_i)}{2 \pi \det^i} dR,
\end{array}
\end{array} \\
\end{equation}
where $\det^i = (\beta_{i+1} - \beta_0)(\alpha_i - \alpha_0) - (\beta_i - \beta_0)(\alpha_{i+1} - \alpha_0)$. Extending these arguments to the case of $\Gamma > 1$ internal points in the polygon positioned at lattice points $(\alpha_0^{\tau}, \beta_0^{\tau})$ for $\tau=1,...,\Gamma$, the monowall has $\Gamma$ independent non-compact moduli, and the junction positions $x_3^i$ asymptotically depend linearly on each non-compact modulus:
\begin{equation}
\begin{array}{ll}
x_3^i - x_{3,0}^i = \sum\limits_{\tau=1}^{\Gamma} m^i_{\tau} R_{\tau}, & \hspace{1.4cm} \phi^i - \phi_0^i = \sum\limits_{\tau=1}^{\Gamma} n^i_{\tau} R_{\tau}, \\[.5cm]
m_{\tau}^i = \frac{\beta_{i+1} - \beta_i}{\det_i^{\tau}}, & \hspace{1.4cm} n_{\tau}^i = -\frac{(\alpha_{i+1} - \alpha_i)}{ 2 \pi \det_i^{\tau}}, \\[.5cm]
\end{array}
\end{equation}
where $\det_i^{\tau} = (\beta_{i+1} - \beta_{0}^{\tau})(\alpha_i - \alpha_{0}^{\tau}) - (\beta_i - \beta_{0}^{\tau})(\alpha_{i+1} - \alpha_{0}^{\tau})$. The same arguments extend to the remaining types of moduli. 

\begin{figure}[htb]
\begin{center}
\centerline{\includegraphics[scale=.35]{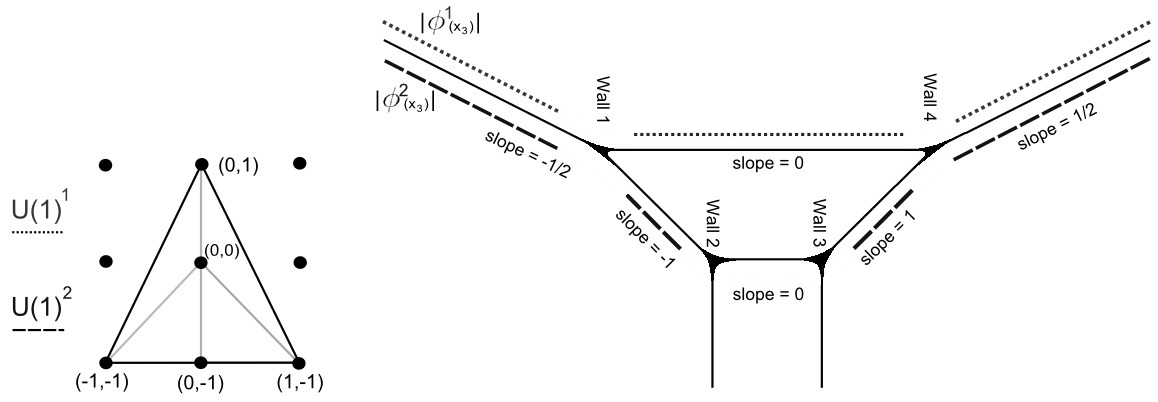}}
\caption{\textit{Left:} Newton polygon (black lines) with a regular triangulation (grey lines). \textit{Right:} Sketch of the amoeba for the associated $U(2)$ monowall, with the two asymptotic Higgs eigenvalues shown in dotted and dashed lines over a range in $x_3$. \label{Amoeba5}}
\end{center}
\end{figure}

The $i^{th}$ subwall has a set of $N$ magnetic charges $g_a^i$ which are determined by the Newton polygon and its triangulation. The charge of a subwall is determined by the difference in slope of the Higgs eigenvalues (which correspond to non-vertical amoeba lines) to either side of the subwall. The magnetic field due to a single, stationary subwall is $\vec{B}_i (x_3) = - \vec{\nabla} \phi_i = g_i \hspace{.1cm} \text{sgn}(x_3 - x_3^i)$. External amoeba lines have slopes $ \left( - \frac{\alpha_{i+1} - \alpha_i}{\beta_{i+1} - \beta_i} \right)$ normal to the corresponding Newton polygon edge and the slopes are triangulation-independent. Internal amoeba lines have slopes $\left( - \frac{\alpha_i - \alpha_{0}^{\tau}}{\beta_{i} - \beta_{0}^{\tau}} \right) $ normal to lines of triangulation and are therefore triangulation-dependent. A subwall which has no effect on the $a^{th}$ eigenvalue has zero charge $g_a^i = 0$ with respect to the $a^{th}$ factor of $U(1)$. A subwall which alters the slope of the $a^{th}$ Higgs eigenvalue $\phi_a(x_3)$ has charge $g_a^i$ equal to half the change in slope. Through the amoeba, the Newton polygon and its triangulation yield precise information about the various asymptotic Higgs eigenvalues $\{ \phi_a(x_3) \}$. The lattice height $N$ of the Newton polygon is the number of $U(1)$ factors from the maximally broken $U(N)$, and each horizontal strip of the lattice is associated with a $U(1)$ factor (see Figure \eqref{Amoeba5}). A subwall whose sub-triangle has lattice height $h$ and occupies $h$ horizontal strips is magnetically charged with respect to each of those $h$ factors of $U(1)$. A Higgs eigenvalue with slope $k/l$ in some region actually represents $l$ degenerate Higgs eigenvalues. To see this illustrated, see Figure \eqref{Amoeba5}. For example, in Figure \eqref{Amoeba5}, the charges for subwall 1 are $g_1^1 = -\frac{1}{4}$ and $g_2^1 = \frac{1}{4}$. For contrast, subwall 2 has charges $g_1^2 = 0$ and $g_2^2 = -\frac{1}{2}$.

Define the vectors $\vec{A} = (\mathcal{A},\mathcal{B},0)$, $d\vec{X} = (-d\mathcal{B}, d\mathcal{A}, dR)$. In terms of the four types of moduli $(\mathcal{A},\mathcal{B},R, \Theta)$ which correspond respectively to $(a,b,x_3,\theta)$, the metric may be written in the Lee-Weinberg-Yi form:
\begin{equation} \label{FinalMetric}
\boxed{ \frac{ds^2}{4 \pi} = \sum\limits_{a=1}^N \hspace{.1cm}\frac{1}{2} U_{ij,a} d\vec{x}^i \cdot d\vec{x}^j + \frac{1}{2} \left[ U^{-1} \right]_{ij,a} 
 \left[ d\theta^i + \vec{W}^{ik}_a \cdot d\vec{x}_k \right] 
 \left[ d\theta^j + \vec{W}^{jl}_a \cdot d\vec{x}_l 
 \right]
 }
\end{equation}
with the following tensors defined:
\begin{align} \label{FinalMetric2}
& \begin{array}{ll}
U_{ii,a} =  g^0_a g_{ia} x_{3i} + \sum\limits^n_{\substack{ j = 1 \\ j \neq i}} g_{ia} g_{ja} \left( \frac{v_a}{G_a} + |x_{3i} - x_{3j} |  \right), 
&  \vec{W}^{ii}_a = \frac{\vec{d}_a}{n} g^i_a - \sum\limits^n_{\substack{ j = 1 \\ j \neq i}} g^i_a g^j_a (\vec{a}^i - \vec{a}^j ), \\[.8cm]
U_{ij,a} = - g_{ia} g_{ja}  \left( \frac{v_a}{G_a} + |x_{3i} - x_{3j} |  \right), 
& \vec{W}^{ij}_a = -\frac{\vec{d}_a}{n} g^j_a + g^i_a g^j_a  (\vec{a}^i - \vec{a}^j ), 
\end{array} \\
& \begin{array}{lll}
\vec{x}^i = \sum\limits_{\tau=1}^{\Gamma} m_{\tau}^i \vec{X}_{\tau} + x^i_{3,0} \hat{x}_3, \hspace{1cm}
& \vec{a}^i = \sum\limits_{\tau=1}^{\Gamma} m_{\tau}^i \vec{A}_{\tau}, \hspace{1cm}
& \theta^i = \sum\limits_{\tau=1}^{\Gamma} m_{\tau}^i \Theta_{\tau}. 
\end{array} \nonumber
\end{align}
Here, $a$ indexes the $N$ $U(1)$ factors. The asymptotic parameters of the metric and the monowall itself, the constant background Higgs, constant background gauge holonomies, the $x_3$ and phase centers of mass, the total magnetic charge, and the slope of the linear background Higgs $\{(v_a, \vec{d}_a, (x_3)_a^{cm}, \vec{a}^{cm}_a, G_a, g^0_a) \}$ relate to the boundary conditions $\{(G_a^{\pm},v_a^{\pm}, \vec{d}_a^{\pm})\}$, which are the left and right charges, Higgs background and holonomy background. The relations are as follows:
\begin{equation} \label{Matching}
\begin{array}{lll}
g^0_a = \frac{1}{2} \left( G^+_a + G^-_a \right), & \hspace{.4cm} v_a = \frac{1}{2} \left( v^+_a + v^-_a \right), & \hspace{.4cm} \vec{d}_a = \frac{1}{2} \left( d^+_a + d^-_a \right), \\
G_a = \frac{1}{2} \left( G^+_a - G^-_a \right), & \hspace{.4cm} x_{3,a}^{cm} = -\frac{1}{2G_a} \left( v^+_a - v^-_a \right), & \hspace{.4cm} \vec{a}_a^{cm} = -\frac{1}{2G_a} \left( d^+_a - d^-_a \right).
\end{array}
\end{equation}

In summary, we have in this section approximated the asymptotic BPS monowall moduli space metric by modeling its asymptotics as abelian interactions of its $n$ well-separated sub-monowalls in a linear Higgs background. Rather than the general $4n$ moduli, consistency conditions reduce the number of moduli to four times the number of internal points in the Newton polygon. We gave the explicit example for one internal point, in which the BPS monowall has but four moduli $(\mathcal{A},\mathcal{B},R,\Theta)$. Each regular Newton polygon triangulation \cite{Gelfand:1994tk} yields a set of subwall magnetic charges and therefore each regular triangulation corresponds to a different sector of the moduli space. With the parameters listed above, the monowall asymptotics in each of the $N$ factors of $U(1)$ are determined by the Higgs spectral curve and Newton polygon. The metric $\eqref{FinalMetric}$ gives the dynamics of well-separated sub-monowalls in terms of the moduli of the monowall.

\section{Conclusions}

A doubly periodic magnetic monopole, known as a monowall, is a magnetic monopole on $\mathbb{R} \times T^2$ with internal phase degrees of freedom whose excitations generate electric charge in the monowall. In $U(N)$ classical Yang-Mills-Higgs gauge theory we employ the Higgs curve, its Newton polygon, and its amoeba to establish the asymptotic behavior of a monowall that has moduli. These tools give us an intuitive picture of the monowall in terms of its constituent charges, or subwalls, and of the symmetries when the charges are widely spread apart. When a modulus of the monowall becomes large, the monowall breaks up into subwalls whose separations vary linearly with respect to the modulus. The subwalls are positioned at locations of partially or fully restored gauge symmetry, a condition that can be inferred from the amoeba. The size of a subwall is the width of the region outside of which gauge and Higgs field interactions can be effectively approximated as abelian. Once the walls are widely-separated with respect to this width, their gauge and Higgs interactions are approximated as $N$ $U(1)$ interactions and emulate classical electromagnetism with a massless Higgs. We proceed to treat the subwalls as uniformly electrically, magnetically and scalar charged planes and write the relativistic Lagrangian, including background gauge and Higgs fields which satisfy a prescription of boundary conditions. For small velocities, this Lagrangian reduces to purely kinetic and we can read off the monowall moduli space metric.

Subwall interactions yield hyperk\"ahler moduli space metrics and hyperk\"ahler \textit{asymptotic} moduli space metrics in the limit that the subwalls are well-separated. The moduli space of a monowall is important in its own right: for small velocities, the subwall dynamics can be approximated as geodesic motion on the moduli space. This moduli space has additional importance to supersymmetric Yang-Mills quantum gauge theory, since moduli spaces of monowalls in Yang-Mills-Higgs theories can be mapped to the Coulomb branch moduli spaces of vacua in the associated five-dimensional quantum field theories.

The asymptotic moduli space of well-separated doubly-periodic monopoles has been addressed previously \cite{Hamanaka:2013lza}, and we expand on this work. In \cite{Hamanaka:2013lza}, the asymptotic monowall moduli space metric was determined for subwalls of identical magnetic charge in $SU(2)$ theory with spatially uniform background fields. 
We generalize from $SU(2)$ theory to $U(N)$ for subwalls of arbitrary magnetic charge and linear background Higgs field, and additionally justify the abelian long-distance approximation by analyzing the Higgs curve and amoeba for large values of a modulus. 
Still, the approach here is limited to well-separated subwalls. While asymptotic moduli spaces have been derived for a variety of monopoles, periodic and non-periodic, the interiors of such moduli spaces remain obscure. The corresponding supersymmetric systems, and in the case of periodic monopoles, the Higgs curve construction may play important roles in future efforts to derive the full moduli space metrics of monopoles.

\textbf{Acknowledgements}

The author is very grateful to Sergey Cherkis for his extensive guidance and helpful exchanges.

\end{document}